# Elastic effects on relaxation volume tensor calculations

B. Puchala,[1,*] M. L. Falk[1], K. Garikipati[2]

[1]Dept. of Materials Science and Engineering, University of Michigan, Ann Arbor, Michigan 48109, USA

[2]Dept. of Mechanical Engineering, University of Michigan, Ann Arbor, Michigan 48109, USA

## ABSTRACT

Relaxation volume tensors quantify the effect of stress on diffusion of crystal defects. Continuum linear elasticity predicts that calculations of these parameters using periodic boundary conditions do not suffer from systematic deviations due to elastic image effects and should be independent of supercell size or symmetry. In practice, however, calculations of formation volume tensors of the <110> interstitial in Stillinger-Weber silicon demonstrate that changes in bonding at the defect affect the elastic moduli and result in system-size dependent relaxation volumes. These vary with the inverse of the system size. Knowing the rate of convergence permits accurate estimates of these quantities from modestly sized calculations. Furthermore, within the continuum linear elasticity assumptions the average stress can be used to estimate the relaxation volume tensor from constant volume calculations.

# I. INTRODUCTION

As semiconductor device dimensions decrease to the nanometer scale, high doping concentrations and very abrupt doping profiles are required to keep resistances low[1]. Precise control of dopant diffusion during processing is necessary, and at the nanometer scale stress effects become increasingly important[2]. Significant and complex stress states can arise from strain-engineering[3,4], lattice and thermal expansion coefficient mismatch, growth stresses, and defect concentrations[5,6]. Stress-affected dopant diffusion has been observed in many systems[6-16], and a thermodynamic formalism has been developed[6,17] relating the stress-affected free energy for the formation and migration of diffusion-mediating defects to the volume changes associated with those processes. Differences between hydrostatic and biaxial stress effects indicate that a full tensorial formulation for the volume change is necessary[6]. Because experimental measurements are difficult, atomistic calculations have found a significant role in determining diffusion mechanisms[18-22], explaining experimental results[23] and parametrizing process models[24].

In purely atomistic calculations, a defect is introduced into material with either free or periodic boundary conditions (FBC or PBC). The atomistic system is allowed to relax and after mechanical equilibrium is reached the resulting energy and volume change are measured. For the finite sizes currently tractable, calculations using FBC can have significant finite-size and surface effects. Calculations using PBC do not have those problems, but instead the periodic supercell creates an infinite array of defects which may interact with each other. It is typically assumed that for calculations using zero average stress PBC, the energy and volume of the supercell in the large supercell limit approach the value for an isolated defect in an infinite medium at zero-pressure. In *ab initio* calculations there are many different factors that influence energy convergence, including the basis-set, brillouin-zone integration, electronic interactions due to supercell shape and size, and inelastic and elastic ionic relaxation[25,26]. Most of these can be dealt with by using more detailed calculations: a larger basis-set, more integration points, or a larger simulation cell to eliminate electronic interactions and inelastic relaxations. Knowledge of the convergence rate can be used to predict values in the infinite cell size limit[27], and correction schemes[28,30] for electrostatic interactions have been proposed to increase the convergence rate.

Elastic effects, particularly related to determining volume changes and therefore stress effects, have received less attention. Potentially significant are (1) elastic image effects from the interaction of the elastic fields of periodically repeated defects, (2) the size and symmetry of the system and the anisotropy of the defect, and (3) changes in the elastic moduli due to bonding changes at the defects. An *ab initio* study by Windl *et al.*[31] found the bulk modulus to converge as the inverse of supercell volume, but within the accuracy of the calculations no effect on the formation volumes was shown. Castleton and Mirbt[27] progressively increased the number of atoms around the defect that were allowed to relax and found that the volume change, defined by the defect's nearest neighbors, converged as the inverse of the volume of the region allowed to relax. By keeping atoms at the boundaries of the supercell fixed elastic image effects were removed, but this prevents the calculation of the thermodynamically relevant volume change of the entire supercell. Probert and Payne[25] suggest relaxing the volume to avoid symmetry effects on structure convergence, and advise that displacement differences between successive shells of atoms be less than some appropriate tolerance before reaching the periodic boundaries to ensure sufficient supercell size.

Our approach is to use continuum linear elasticity to quantify the effects of periodic images, supercell size and symmetry and to use atomistic calculations employing the Stillinger-Weber empirical potential for silicon to consider the bonding effects on the elastic moduli. An empirical potential does not calculate defect parameters as accurately as *ab intio* methods, but it is useful for studying elastic effects because there are no electrostatic interactions, and it allows us to simulate large systems approaching the far-field. In a previous paper[32], we used symmetry to show that, according to linear elasticity, PBC at zero average stress do not affect the calculated relaxation volume tensor of an isotropic defect and demonstrated this by calculating the formation energy and volume of a vacancy in Stillinger-Weber silicon. Here we formally prove that, according to linear elasticity, there are no elastic image effects on the calculated relaxation volume tensor for any anisotropic defect in any shape or size supercell. When the supercell is not allowed to fully relax, linear elasticity shows that the average stress can be used to calculate the relaxation volume tensor. We verify linear elasticity's predictions in the far-field of atomistic calculations by calculating the formation energy and volume tensor for the <110> interstitial in Stillinger-Weber silicon. Finally, we show that in atomistic calculations bonding changes affect the convergence of elastic moduli with system size. Therefore, the relaxation volume tensor is not indepedent of system size as predicted by linear elasticity, but rather converges as the system size increases.

## II. CONTINUUM LINEAR ELASTICITY CALCULATIONS

The dependence of the Gibbs free energy needed for formation or migration of a defect, $G^{f,m}$, as a function of stress, $\sigma_{kl}$, is defined as the formation or migration volume[6]

$$-\frac{\partial G^{f,m}}{\partial \sigma_{kl}} \equiv V_{kl}^{f,m}. \qquad (1)$$

In the case that the defect is a vacancy or self-interstitial, the formation volume is the total change in the system's volume when one of the internal atoms is removed from the bulk to the surface or vice versa and thus

$$V_{kl}^{f} = V_{kl}^{r} \pm \frac{1}{3}\Omega\delta_{kl}, \qquad (2)$$

where $V_{kl}^{r}$ is the relaxation volume, (+/-) is for a (vacancy/interstitial) defect, $\Omega$ is the atomic volume, $\delta_{kl}$ is the Kronecker delta. The last term follows from an assumption that it is equally likely for the defect to form from any surface.

In order to determine the predicted continuum linear elastic effect of PBC on atomistically calculated relaxation volumes of anisotropic defects, we first review our previous derivation which relates the observed volume change to the surface tractions. We use a slight notation change in comparison with our earlier work[32] to clarify the distinction between the observed volume change and the boundary condition independent relaxation volume tensor. We model the point defect as a center of expansion or contraction via a force dipole, $D_{ij}$ in a finite elastic body, $B$. The equilibrium equation for the center of expansion or contraction is

$$\frac{\partial \sigma_{jm}}{\partial x_m} + D_{jm}\frac{\partial \delta(x-x')}{\partial x_m} = 0. \qquad (3)$$

The dipole causes a strain field, $\varepsilon_{kl}$, and for any boundary conditions the observed tensorial volume change from the original defect-free state, $\Delta V_{kl}$, is

$$\Delta V_{kl} = \int_{B} \varepsilon_{kl} dV. \qquad (4)$$

From the stress-strain relation this is

$$\Delta V_{kl} = S_{klij} \int_B \sigma_{ij} dV ,$$  (5)

where $S_{klij}$ is the constant compliance tensor. Note that Eq. (5) is not directly comparable to atomistic calculations because $\sigma_{ij}$ is the continuum stress field and contains a singularity, which does not exist in atomistic calculations. To evaluate the effect of boundary conditions we wish to use the divergence theorem to relate this to a surface integral. Therefore we write Eq. (5) as

$$\Delta V_{kl} = S_{klij} \int_B \left[ \frac{\partial(x_i \sigma_{mj})}{\partial x_m} - x_i \frac{\partial \sigma_{mj}}{\partial x_m} \right] dV .$$  (6)

It will become useful to divide the domain into a region $D$ enclosing the center of expansion or contraction, and a region surrounding it, $B - D$, so we write

$$\Delta V_{kl} = S_{klij} \left\{ \int_D \left[ \frac{\partial(x_i \sigma_{mj})}{\partial x_m} - x_i \frac{\partial \sigma_{mj}}{\partial x_m} \right] dV + \int_{B-D} \left[ \frac{\partial(x_i \sigma_{mj})}{\partial x_m} - x_i \frac{\partial \sigma_{mj}}{\partial x_m} \right] dV \right\} .$$

Substituting in Eq. (3) gives

$$\Delta V_{kl} = S_{klij} \left\{ -\int_D x_i D_{jm} \frac{\partial \delta(x - x')}{\partial x_m} dV + \int_B \frac{\partial(x_i \sigma_{mj})}{\partial x_m} dV - \int_{B-D} x_i \frac{\partial \sigma_{mj}}{\partial x_m} dV \right\} .$$  (7)

We evaluate the first term using the standard result for the spatial derivative of the Dirac-delta function, and the second term using the symmetry of the stress tensor and the divergence theorem. The third terms drops out due to the fact that $\sigma_{mj}$ is divergence-free over $B - D$. Using these three results in Eq. (7) we obtain a modified version of our previous result,

$$\Delta V_{kl} = -S_{klij} D_{ij} + S_{klij} \int_{Surface} x_i \sigma_{jm} n_m dA .$$  (8)

With FBC, the surface integral vanishes and the volume change defines the relaxation volume tensor

$$V_{kl}^r \equiv -S_{klij} D_{ij} . \tag{9}$$

The difference $\Delta V_{kl}$ between FBC and arbitrary boundary conditions can be determined by evaluating the integral in Eq. (8). However, in atomistic calculations with PBC it is more convenient to compare with a volume integral. Therefore we rewrite the second term of (8) using a stress field $\overline{\sigma}_{jm}$, such that $\overline{\sigma}_{jm} = \sigma_{jm}$ in $B - D$, and $\overline{\sigma}_{jm}$ is non-singular and divergence-free in $D$. We then use the fact that $\overline{\sigma}_{jm}$ so-defined is divergence-free over all of $B$ to obtain the expression

$$\Delta V_{kl} = V_{kl}^r + S_{klij} \int_B \overline{\sigma}_{kl} dV . \tag{10}$$

The field $\overline{\sigma}_{kl}$ is a good model for an atomistic stress field since it is non-singular and divergence-free over $B$ and is equal to the elastic field in $B - D$. In what follows we will use the virial formulation for $\overline{\sigma}_{kl}$.

To compare Eq. (10) to atomistic calculations with PBC we note that supercell relaxation occurs by changing the magnitude and direction of the vectors that define the supercell. The change in these vectors defines an average strain relative to the perfect structure, so we find

$$\Delta V_{kl} = V^{ref} \varepsilon_{kl}^{avg} = V_{kl}^r + S_{klij} \sigma_{ij}^{avg} V^{ref} . \tag{11}$$

The stress $\sigma_{ij}^{avg}$ is the average over $B$, as would be measured using the virial formulation, and since Eq. (11) is derived using the assumptions of linear elasticity theory the integral is carried out over the entire undeformed volume, $V^{ref}$. This result shows that there are no elastic image effects, and it holds for defects of any anisotropy, and supercells of any shape or size. The observed volume change in atomistic calculations is the relaxation volume plus a correction term that arises if the system is not allowed to fully relax. For PBC with zero average stress, linear elasticity predicts that the observed volume change is exactly the relaxation volume. For self-equilibrated stress states in paralleliped-shaped supercells the average surface stress must equal the resolved virial stress, so Eq. (11) is equally valid if $\sigma_{ij}^{avg}$ is measured from the forces crossing the supercell boundaries. Using this fact, Eq. (11) can also be derived for PBC by evaluating the surface integral in Eq. (8). We include the derivation in Appendix A.

Given arbitrary boundary conditions we also want to calculate the effect on defect formation energies. From the continuum elasticity point of view, a constant volume calculation can be viewed as a two-step process. The first step consists of inserting a defect into a supercell and allowing it to relax to zero mean stress. The second step is a transformation that returns the supercell to the constant volume shape and size. The relaxed supercell and the constant volume supercell are both parallelepipeds, so the stress/strain field, $\sigma_{ij}^{CV} = S_{ijkl}\varepsilon_{kl}^{CV}$, that performs the return transformation is uniform. Since the average stress in the relaxed supercell is zero, the average stress measured in the constant volume supercell is $\sigma_{ij}^{avg} = \sigma_{ij}^{CV}$. In order to determine the zero-stress formation energy of the defect, $E^f$, we must subtract the work done on the body in the return transformation from the formation energy found in the constant volume supercell

$$E^f = E^{f,CV} - \int_B \frac{1}{2}\sigma_{ij}^{CV}\varepsilon_{ij}^{CV}dV$$

$$E^f = E^{f,CV} - V^{ref}\frac{1}{2}\sigma_{ij}^{avg}S_{ijkl}\sigma_{kl}^{avg}. \qquad (12)$$

This gives the finite-crystal strain energy. As we showed previously[32] the finite crystal strain energy converges to the infinite-crystal strain energy as $E^{\infty} - E^{finite} \propto N^{-1}$, $N$ being the number of atoms.

This results of this section are a validation that the supercell approach does not introduce any systematic errors due to elasticity. In the following section we perform atomistic calculations to show that in practice system size does effect the observed volume change. FBC introduce surfaces and PBC an infinite array of defects that change the elastic moduli and result in deviations from elasticity's prediction at small cell sizes.

## III. ATOMISTIC CALCULATIONS

We calculated the formation energy and volume tensor of a <110> dumbbell interstitial in Stillinger-Weber silicon by energy minimization using the conjugate gradient method. The Stillinger-Weber potential is a commonly used empirical potential for silicon[33,34], and as such it is not as accurate as *ab initio* calculations near the defect, but is useful for our purposes since the decreased computational costs allow us to use the large system sizes necessary to check the predictions of continuum linear elasticity in the far-field.

### A. Methods

We calculated formation energies and volumes for cubic systems ranging in size from 64 to 110,592 atoms. The <110> dumbbell interstitial was constructed by displacing an atom near the center of the simulation cell by (-0.162, -0.162, +0.1325) unit cells and adding an interstitial that is displaced (+0.162, +0.162, +0.1325) from the first atom's original position. Upon relaxation the atoms composing the dumbbell relax in the z-direction away from the neighboring atoms in the <110> chain, so to speed convergence the dumbbell atoms were given the initial z-displacement indicated above.  The sign of the z-displacement depends on which atomic site the dumbbell is located.

The <110> dumbbell interstitial in Stillinger-Weber silicon was found to have two different configurations with nearly equal formation energy, shown in Fig. 1.  The major difference between the two is that the lower energy configuration, which we call (A), had non-zero $V_{xz}^f$ and $V_{yz}^f$, while for the higher energy configuration, which we call (B), $V_{xz}^f$ and $V_{yz}^f$ are zero.  As can be seen in Fig. 1, the non-zero $V_{xz}^f$ and $V_{yz}^f$ is manifested locally by the dumbbell tilting and breaking the symmetry about the $(110)$ plane.  Both structures maintain symmetry about $(1\bar{1}0)$.  For both FBC and PBC an initially perfect crystal with a defect, small random displacements, and less than 512 atoms relaxed to (A), with 512 atoms the crystal relaxed to (A) or (B), and with greater than 512 atoms the crystal became stuck in configuration (B).  We attempted several schemes of increasing complexity to ensure minimization to the lower energy configuation (A) at large system sizes, and the successful method involved taking a relaxed (A) configuration at one system size and adding atoms at the surface to construct the next largest system size.  The new atoms were positioned according to the final average strain state of the previous system.

Three boundary conditions were used:  FBC, PBC at constant pressure (PBC CP), and PBC at constant volume (PBC CV) equal to the volume of the relaxed defect-free system.  We check that there are no elastic image effects by comparing the FBC and PBC calculations, and check the second term in Eq. (11) and Eq. (12) by comparing PBC CP and PBC CV calculations. We also imposed small random initial displacements of approximately 1% of the atomic spacing on all the atoms and created ten samples for each system size. The energy of the system was then minimized from this starting configuration using the conjugate gradient method. The minimization was considered complete when fourteen sequential iterations each resulted in less than a 1 neV reduction in energy, with the last seven sequential iterations also producing less than a $10^{-4}$ Å$^3$ change in volume.

As noted in the previous section, at mechanical equilibrium with the parallelepiped-shaped computational cells used, a relaxation to zero surface traction is identical to relaxation to zero average volumetric stress as measured by the virial formulation. However, in practice we found it preferable to use the virial stress for two reasons. First, in diamond cubic silicon before mechanical equilibrium is reached, xy-shearing results in a state of alternating positive and negative stress between (001) planes as the interpenetrating FCC lattices attempt to relax internally relative to each other. If the stress is only calculated at a single boundary, this gives an inaccurate measurement of the overall stress state and impedes convergence to zero average stress. Second, given the same nominal stress convergence criteria, the virial stress is stricter because it averages over the entire cell while the boundary stress only averages over the boundary. At cell sizes from 64 up to 21,952 atoms the energies and volumes measured using the zero virial stress condition matched the energies and volumes resulting from using the zero average surface traction condition in the periodic case or relaxation in the absence of constraints for free surface boundary conditions. However, the spread in values was reduced when the zero virial stress condition was employed. For these reasons we used the virial formulation to calculate an average stress tensor in the computational cell and zero average stress ($\pm 10^{-2}$ Pa) was maintained by scaling atomic positions and, if present, periodic boundaries. The elastic moduli of Stillinger-Weber silicon were used to adjust the strain on the system in order to maintain zero average stress during the relaxation process. In the FBC case, after the energy was minimized in this way, rescaling toward the zero average stress condition was discontinued and the energy was again minimized to reach a zero surface traction condition. We found that this method reduced scatter in the formation volume of FBC samples.

### B. Measurements

As discussed previously, formation volume measurements are straightforward for periodic boundary conditions. Strain is defined by the position of the periodic boundaries, and each component of the relaxation volume is determined by multiplying the corresponding strain component by the perfect reference volume

$$V_{ij}^r = \varepsilon_{ij}^{avg} V^{ref} \,. \tag{13}$$

Then the formation volume is

$$V_{ij}^f = \varepsilon_{ij}^{avg} V^{ref} \pm \frac{1}{3}\Omega \delta_{ij} \,. \tag{14}$$

For FBC, the volume change of an elastic body must be determined from the displacement of the surface of the sample, according to[32]

$$V_{ij}^f = \int\limits_{Surface} \frac{1}{2}\left(u_i n_j + u_j n_i\right) dA \pm \frac{1}{3}\Omega \delta_{ij} \,, \tag{15}$$

where $u$ is the displacement and $n$ is the surface normal. In an atomistic simulation this is a finite sum of individual atomic displacements and the differential area is the average surface area per surface atom. This method is not appropriate for PBC because it does not take into account strain between atoms on either side of the periodic boundary, which is small, but significant when multiplied over the area of the boundary.

We calculated elastic moduli for defect-free Stillinger-Weber silicon by using a single perfect unit cell with periodic boundaries. The bulk modulus, $K$, was calculated by measuring the volume change under hydrostatic pressure of $\sigma_h = -100$ MPa and $\sigma_h = 100$ MPa. Then

$$K = \frac{\Delta \sigma_h}{\Delta V} V(\sigma = 0) \ . \qquad (16)$$

Additionally, $\sigma_{xy}$ was measured for $\varepsilon_{xy} = \varepsilon_{yx} = 0.001$ so that $C_{44}$ could be calculated

$$C_{44} = \frac{\sigma_{xy}}{2\varepsilon_{xy}} . \qquad (17)$$

The diamond cubic structure can be thought of as two interpenetrating FCC lattices, which can relax internally and produce an internal strain. We measured $C_{44}$ with internal strain because it does occur after a defect is introduced. We also calculated the bulk and shear moduli as a function of system size for both FBC and PBC systems with an interstitial. These systems were created for $N = 216$ to $110,592$ atoms without any random initial displacements and tested as above, except stress control had to be used for the FBC case. In this case the strains were calculated using a relaxed system with an intersitial and the same number of atoms as a reference state.

### C.  Atomistic Results

As predicted by continuum linear elasticity, the formation energies and volume tensors calculated with FBC and PBC, at both constant pressure and volume, converge to the same value in the large-size limit, as shown in Fig. 2 and 3. The 64 atom FBC samples underwent surface reconstructions and are not included in the plots or analysis. No other samples underwent surface reconstructions, and in the Stillinger-Weber potential the equilibrium distance and angle is not coordination dependent, so surface stress is not a factor in these results.

The formation energy converges more rapidly with PBC than FBC, as seen in Fig. 2. As predicted by continuum linear elasticity, and shown in Fig. 4, formation energy converges in the large-size limit, with the error decreasing as $1/N$,

$$\log\left(E^f(N) - E^{f,\infty}\right) = const - \log(N). \tag{18}$$

The converged formation energies, $E^{f,\infty}$, were determined from the formation energies at a given system size, $E^f(N)$, by fitting the data to Eq. (18). The converged energies are $E^{f,\infty}_{Sil,<110>,(A)}$ = 4.7091 eV, and $E^{f,\infty}_{Sil,<110>,(B)}$ = 4.7122 eV, with uncertainty no greater than $10^{-4}$ eV.

The formation energies are in agreement with the values calculated elsewhere for the Stillinger-Weber potential[34-37]. Some of the literature seems to confuse the <110> dumbbell and what is generally called the "extended" interstitial. The extended interstitial is lower energy than the <110> dumbbell in empirical calculations[35,38], but was found to be metastable in an *ab initio* calculation[39]. No other Stillinger-Weber results are known for the full <110> formation volume tensor.

Contrary to the prediction of continuum linear elasticity, the formation volume tensor was also found to converge with system size, the error decreasing as $1/N$, as shown in Fig. 5. The trace of the formation volume tensor converges much more rapidly with PBC than FBC. The convergence of each component of the volume tensor is not shown but similar. The converged values of the formation volume tensor were determined similarly to the formation energies, and are

$$\overline{\overline{V}}_{SiI,<110>,(A)}^{f,\infty} = \begin{bmatrix} 9.585 & 7.972 & \pm 2.176 \\ 7.972 & 9.585 & \pm 2.176 \\ \pm 2.176 & \pm 2.176 & -5.493 \end{bmatrix} \text{Å}^3 \text{, and}$$

$$\overline{\overline{V}}_{SiI,<110>,(B)}^{f,\infty} = \begin{bmatrix} 8.871 & 7.296 & 0 \\ 7.296 & 8.871 & 0 \\ 0 & 0 & -4.849 \end{bmatrix} \text{Å}^3 \text{,}$$

with uncertainty in the components of $\overline{V}^{f,\infty}$ no greater than $10^{-3}$ Å$^3$. The (±) for (A) indicates that it is physically equivalent for the tilt to be in either direction since the sign of shear strains is arbitrary.

The principle axes of the <110> (A) formation volume tensor are tilted 7.5° off the (001) plane. To our knowledge, this is the first report of the tilted <110> interstitial. This may be due to a focus on the energy of the defect rather than on the structure of the relaxation in previous Stillinger-Weber studies. In a recent *ab initio* study[40] there was not any tilting in the <110> interstitial, despite allowing the full relaxation of the supercell, indicating that the tilt is likely to be an artifact of the Stillinger-Weber potential.

Continuum linear elasticity's inability to predict the formation volume tensor convergence with system size is due to its assumption that the elastic moduli are constant. The slow convergence of the formation volume tensor with FBC is caused by the slow convergence of elastic moduli, as shown in Fig. 5. Decreased coordination of surface atoms results in decreased stiffness. The similarity to the formation volume convergence is apparent. Note that Fig. 5 plots only the absolute values of the convergence, therefore the direction of the convergence can not be determined from the figure. We observed that in systems with FBC the elastic moduli increase with system size, matching the observed decrease with system size in the magnitude of the volume relaxation. In other words, as the moduli increase with system size the outward relaxation around the interstitial decreases in order to reduce the strain energy in the surrounding system. In systems with PBC, due to bonding changes at the defect there is a slight decrease in the bulk modulus and a slight increase in the shear modulus with increasing system size. The trace of the formation volume tensor shows an increase in the magnitude of the relaxation around the interstitial with increasing system size, corresponding to the decrease in the bulk modulus. Thus, the convergence of the elastic moduli in the system with a defect to the elastic moduli of the defect-free system is an indication of the convergence of the formation volume tensor and a $1/N$ form for the error can be expected. The elastic moduli convergence is in agreement with the results of Windl $et\ al.$[31] and suggests that either the precision of those calculations or electrostatic effects are hiding the associated formation volume convergence. Similar to the approach of Castleton and Mirbt[27], we can use the convergence rate to estimate the final converged formation volume tensor. At this convergence rate, with formation volume calculations for systems with $N_1$ and $N_2$ atoms, the estimated formation volume in the large-size limit is

$$V^{f,\infty}(N_2) = \frac{N_1 V^f(N_1) - N_2 V^f(N_2)}{N_1 - N_2}. \tag{19}$$

Figure 6 shows how this estimate converges with system size, allowing us to estimate the converged formation volume within the accuracy of the measurements, $10^{-3}$ Å$^3$, by extrapolating from the $N_1 = 512$ and $N_2 = 1728$ systems.

Finally, the agreement between calculations with PBC CP and PBC CV in Fig. 2 and 3 demonstrates that correction terms calculated using linear elasticity do hold, at least to a good approximation. When the system is not allowed to fully relax we can adjust the formation volume tensor by using Eq. (11), and we can adjust the formation energy to account for elastic strain energy by using Eq. (12). At small system sizes, Fig. 2 and 3 show that there are small differences between PBC CP and PBC CV which can either be attributed to using the elastic moduli of a perfect system in Eqs. (11) and (12), rather than the actual moduli of the system with a defect, or to strain dependence of the moduli. As the system size increases both effects decrease and the calculations converge in the large-size limit.

## IV. CONCLUSIONS

We have validated the supercell approach for calculating relaxation volume tensors by formally showing that, according to linear elasticity, the calculated relaxation volume tensor of any anisotropic defect in any shape or size supercell is not affected by PBC at zero average stress. This rigorously demonstrates why the supercell approach can provide accurate calculations of relaxation volume tensors. When the supercell is not allowed to fully relax, the average stress can be used to calculate the relaxation volume tensor. Atomistic calculations verify linear elasticity's predictions in the far-field for an anisotropic <110> interstitial in Stillinger-Weber silicon, and show that, in practice, bonding changes at the defect result in elastic moduli changes. The observed $1/N$ decrease of the error of the relaxation volume tensor is due to to the convergence of the elastic moduli. Knowledge of this convergence rate allows for accurate estimation of the relaxation volume tensor with relatively modest simulation sizes.

## ACKNOWLEDGMENTS

This research was supported by the US National Science Foundation through grant number CMS0331016.  This support is gratefully acknowledged.

## APPENDIX A

As a more complicated alternative, we can also derive Eq. (11) for PBC from Eq. (8) by evaluating the surface integral, as follows. Consider a periodic array of simulation cells, each enclosing a point defect of arbitrary anisotropy, and defined by the lattice vectors $v_1$, $v_2$, and $v_3$ which are defined in the Cartesian coordinate system. It is common to perform atomistic simulations with periodic supercells defined by a coordinate system that is not orthogonal, typically BCC or FCC systems, so we do not restrict the lattice vectors to be orthogonal. Let $\Gamma_k^l$ and $\Gamma_k^r$ be a pair of boundaries with area $A_k$, parallel to each other, and bounding one cell. For example, $A_1$ is the bounding area defined by $v_2$ and $v_3$. Let $n_k$ be the unit outward normal vector to $\Gamma_k^l$ and $\Gamma_k^r$. Traction continuity at the periodic boundaries gives the condition $\sigma_{jm}^+ n_m^+ = -\sigma_{jm}^- n_m^-$, where $(\bullet)^\pm$ denotes the limiting values of a quantity as $\Gamma$ is approached from either side. The surface integral in Eq. (3b) becomes

$$\int_{Surface} x_i \sigma_{jm} n_m \, dA = \sum_{k=1}^{3} \int_{\Gamma_k^l \cup \Gamma_k^r} x_i \sigma_{jm} n_m \, dA \tag{A1}$$

$$= \sum_{k=1}^{3} \left( \int_{\Gamma_k^l} x_i^- \sigma_{jm}^- n_m^- \, dA + \int_{\Gamma_k^r} x_i^+ \sigma_{jm}^+ n_m^+ \, dA \right). \tag{A2}$$

By $n_m^+ = -n_m^-$ and traction continuity this is

$$\int_{Surface} x_i \sigma_{jm} n_m \, dA = \sum_{k=1}^{3} \int_{\Gamma_k^r} \left( x_i^+ \sigma_{jm}^+ n_m^+ - x_i^- \sigma_{jm}^+ n_m^+ \right) dA \tag{A3}$$

$$= \sum_{k=1}^{3} \int_{\Gamma_k^r} \left( x_i^+ - x_i^- \right) \sigma_{jm}^+ n_m^+ \, dA. \tag{A4}$$

The quantity $\left(x_i^+ - x_i^-\right)$, the i$^{th}$ component of distance between positions with traction continuity on the opposite boundaries is simply the i$^{th}$ component of $v_k$, giving

$$\int_{Surface} x_i \sigma_{jm} n_m dA = \sum_{k=1}^{3} \int_{\Gamma_k'} v_{ki} \sigma_{jm} n_{km} dA_k \, . \qquad (A5)$$

The quantities $v_k$ and $n_k$ are constants, so the integrals are equal to the average stress on the boundary multiplied by the area. In mechanical equilibrium the average traction on each face is equal to the resolved volumetric mean stress

$$= \sum_{k=1}^{3} \left( v_{ki} \left( \sum_{m=1}^{3} \sigma_{jm}^{avg} n_{km} \right) A_k \right) \qquad (A6)$$

$$= \sum_{m=1}^{3} \left( \sigma_{jm}^{avg} \left( \sum_{k=1}^{3} A_k v_{ki} n_{km} \right) \right). \qquad (A7)$$

Which by symmetry of the stress tensor and the identity shown in Appendix B is

$$= V \sigma_{ij}^{avg} \, . \qquad (A8)$$

As before we note that Eq. (8) is derived using the assumptions of linear elasticity theory. This means that the integral is carried out over the surface of the undeformed volume and thus the volume in Eq. (A8) is the undeformed, reference volume. Therefore

$$\int_{Surface} x_i \sigma_{jm} n_m dA = V^{ref} \sigma_{ij}^{avg} \, , \qquad (A9)$$

and combining Eqs. (8) and (A9) we obtain Eq. (11).

Since $A_1 n_1 = v_2 \times v_3$, $A_2 n_2 = v_3 \times v_1$, and $A_3 n_3 = v_1 \times v_2$,

$$\sum_{k=1}^{3} A_k v_{ki} n_{km} = v_1 \otimes (v_2 \times v_3) + v_2 \otimes (v_3 \times v_1) + v_3 \otimes (v_1 \times v_2) \qquad \text{(B1)}$$

With Eq. (B1) and the following theorem, we get the result in the text.

**Theorem**: If $a, b, c$ are three linearly independent, but otherwise arbitrary vectors, then the matrix $A = a \otimes (b \times c) + b \otimes (c \times a) + c \otimes (a \times b)$ satisfies $A = a \cdot (b \times c) I$, where $I$ is the $3 \times 3$ identity matrix.

**Proof**: Let $a^* = \dfrac{a}{|a|}, b^* = \dfrac{b}{|b|}, c^* = \dfrac{c}{|c|}$. Note that

$$a^* \cdot A a^* = a^* \cdot (a \otimes (b \times c) + b \otimes (c \times a) + c \otimes (a \times b)) a^* = (a \cdot a^*)(b \times c) \cdot a^* = a \cdot (b \times c) \qquad \text{(B2a)}$$

Likewise,

$$b^* \cdot A b^* = a \cdot (b \times c). \qquad \text{(B2b)}$$

$$c^* \cdot A c^* = a \cdot (b \times c) \qquad \text{(B2c)}$$

Also,

$$a^* \cdot A b^* = b^* \cdot A a^* = (a \cdot (b \times c))(a^* \cdot b^*), \qquad \text{(B3a)}$$

$$b^* \cdot A c^* = c^* \cdot A b^* = (a \cdot (b \times c))(b^* \cdot c^*), \qquad \text{(B3b)}$$

$$c^* \cdot A a^* = a^* \cdot A c^* = (a \cdot (b \times c))(c^* \cdot a^*). \qquad \text{(B3c)}$$

Now, since any vector $u$ has a unique decomposition as $u = u_a a^* + u_b b^* + u_c c^*$, it follows from direct substitution of (2a—2c) and (3a—3c) that $u \cdot (A - a \cdot (b \times c)I)u = 0$. Since $u$ is arbitrary the result follows.

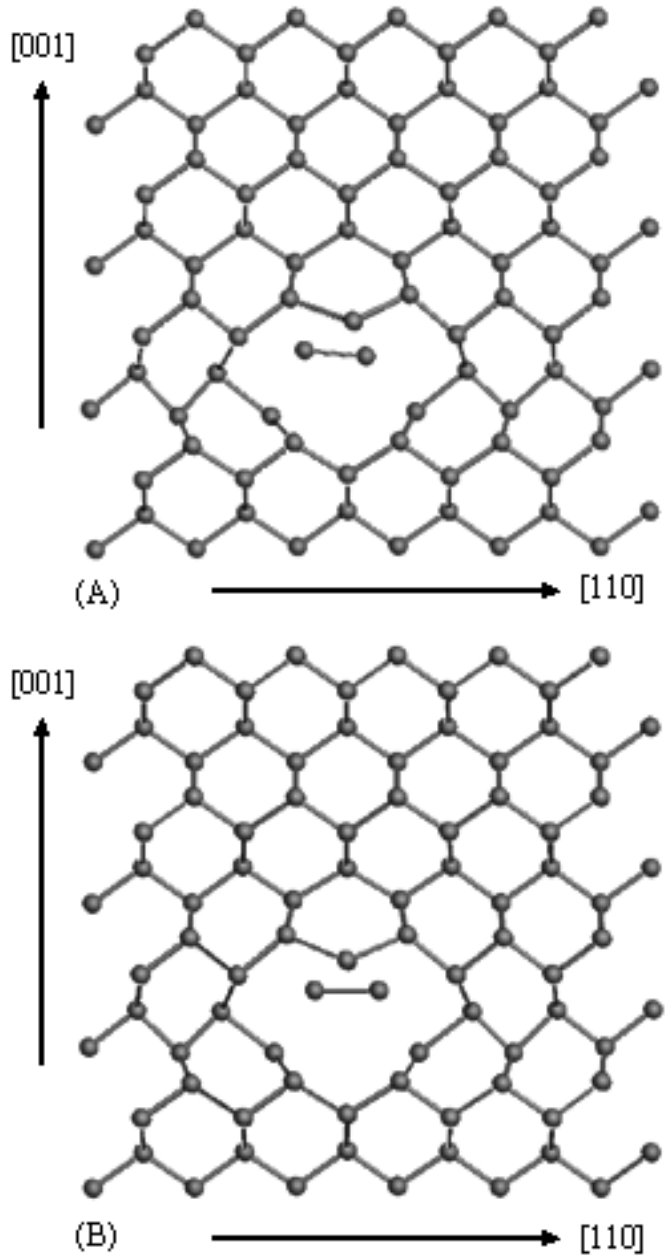

FIG. 1. Structure of the relaxed Si <110> dumbbell interstitial with displacements scaled by 3x for clarity. In configuration (A), the dipole tilts and breaks the symmetry about the $(110)$ plane, resulting in a slightly lower energy than in configuration (B). Both (A) and (B) maintain symmetry about the $(1\bar{1}0)$ plane.

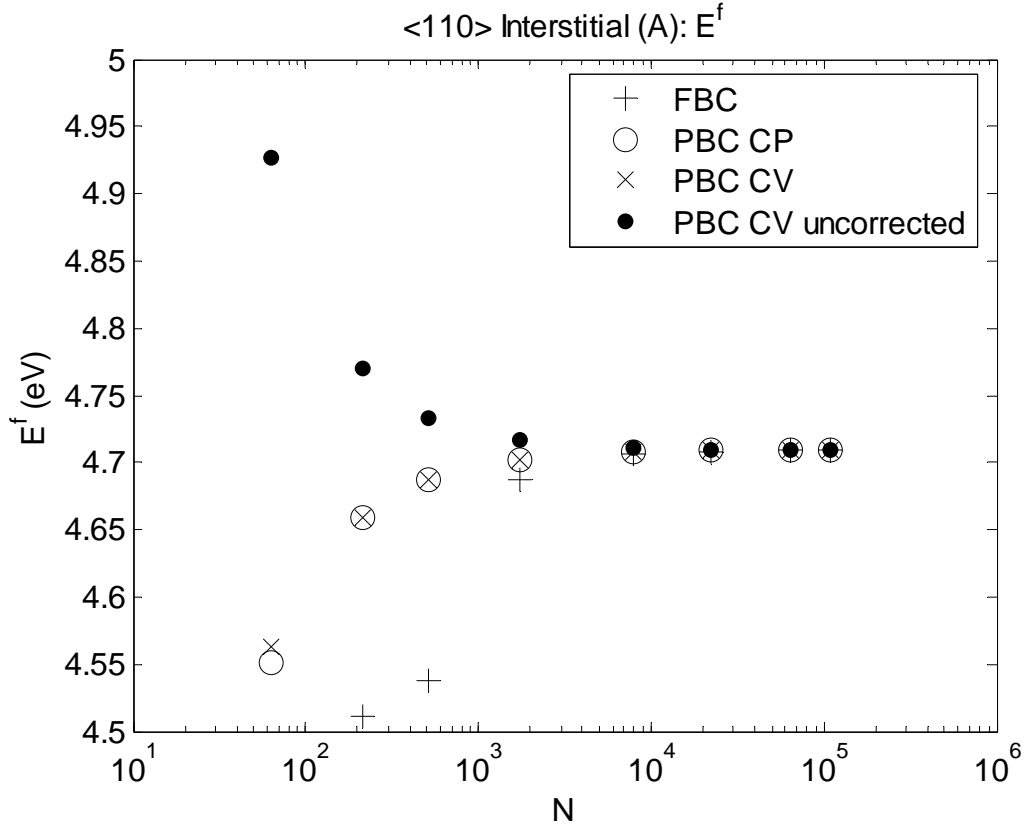

FIG. 2. In the large-size limit the formation energy of an isolated <110> interstitial (A) in Stillinger-Weber silicon converges to $E_{SiI,<110>,(A)}^{f,\infty} = 4.7091$ eV for all measurements: free boundary conditions (FBC); periodic boundary conditions at constant pressure (PBC); periodic boundary conditions at constant volume equal to the reference volume (PBC CV uncorrected); periodic boundary conditions at constant volume equal to the reference volume with the energy correction, Eq. (12) for elastic stress (PBC CV).

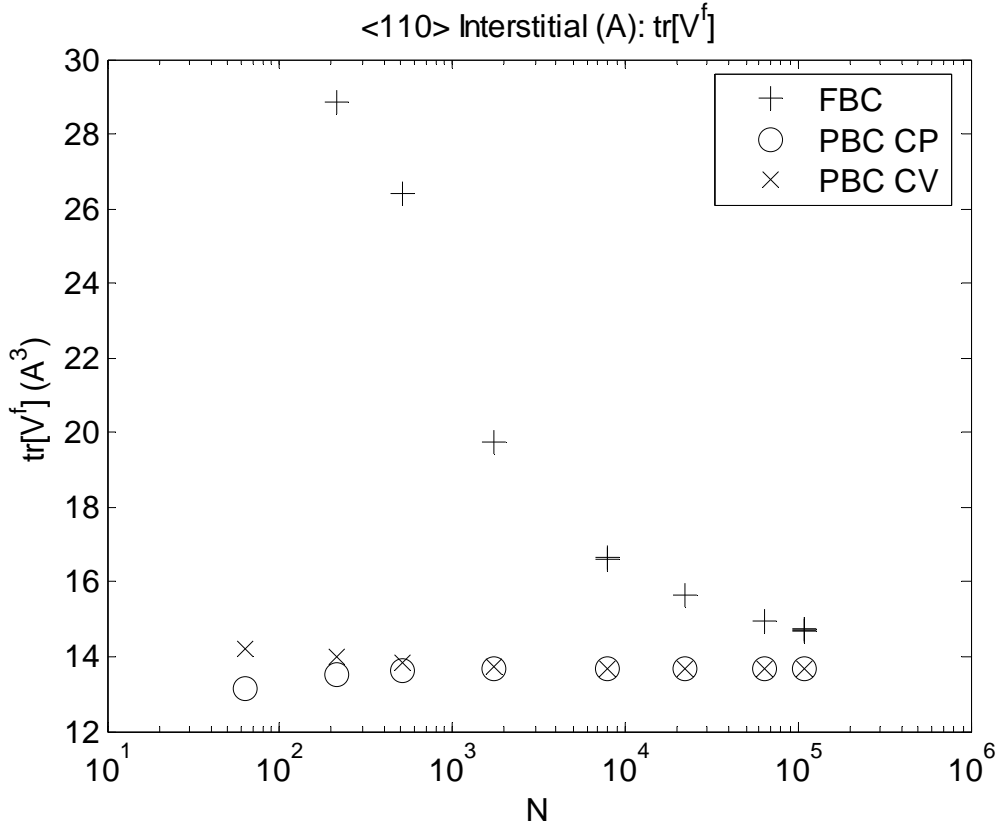

FIG. 3. In the large-size limit the trace of the formation volume of an isolated vacancy in Stillinger-Weber silicon converges to $\mathrm{tr}[\overline{\overline{V}}_{Si,<110>,(A)}^{f,\infty}] = 13.677$ Å$^3$ for all boundary conditions: free boundary conditions (FBC); periodic boundary conditions at constant pressure (PBC); periodic boundary conditions at constant volume equal to the reference volume with the volume correction, Eq. (11) for elastic stress (CV).

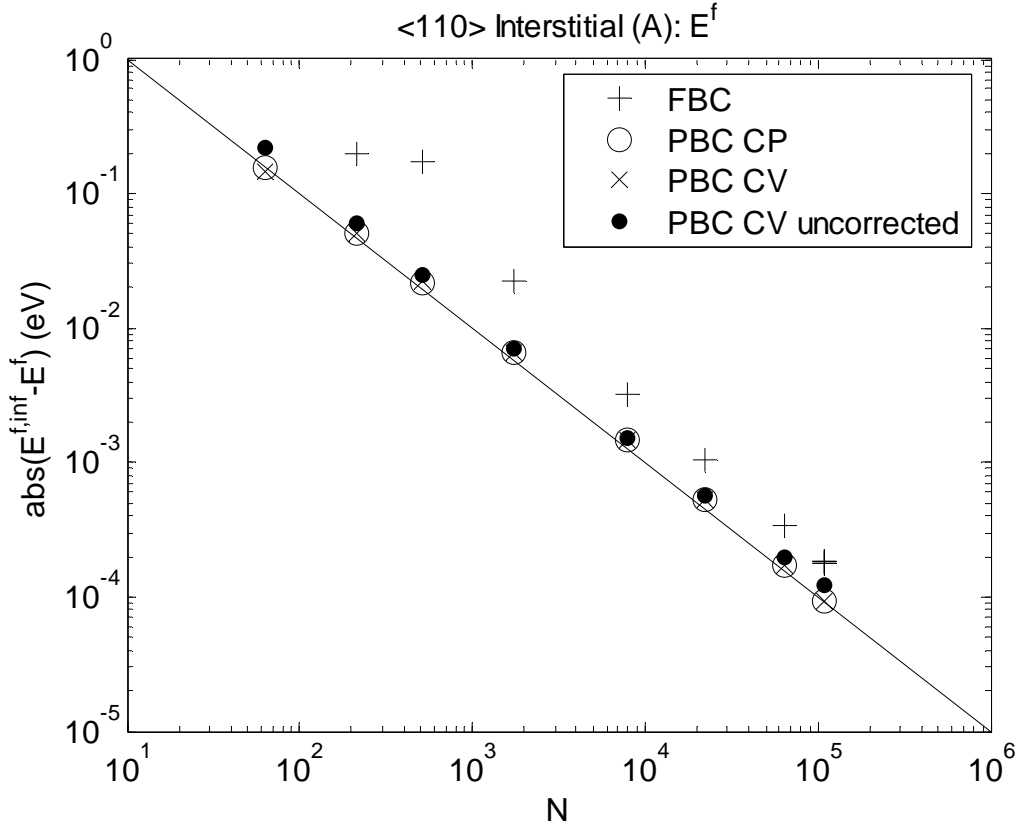

FIG. 4. The formation energy converges in the large-size limit to $E^{f,\infty}_{Sil,<110>,(A)} = 4.7091$ eV, with the error decreasing as the inverse of the system size $N$, for all boundary conditions: free boundary conditions (FBC); periodic boundary conditions at constant pressure (PBC); periodic boundary conditions at constant volume equal to the reference volume (PBC CV uncorrected); periodic boundary conditions at constant volume equal to the reference volume with the energy correction, Eq. (12) for elastic stress (PBC CV).

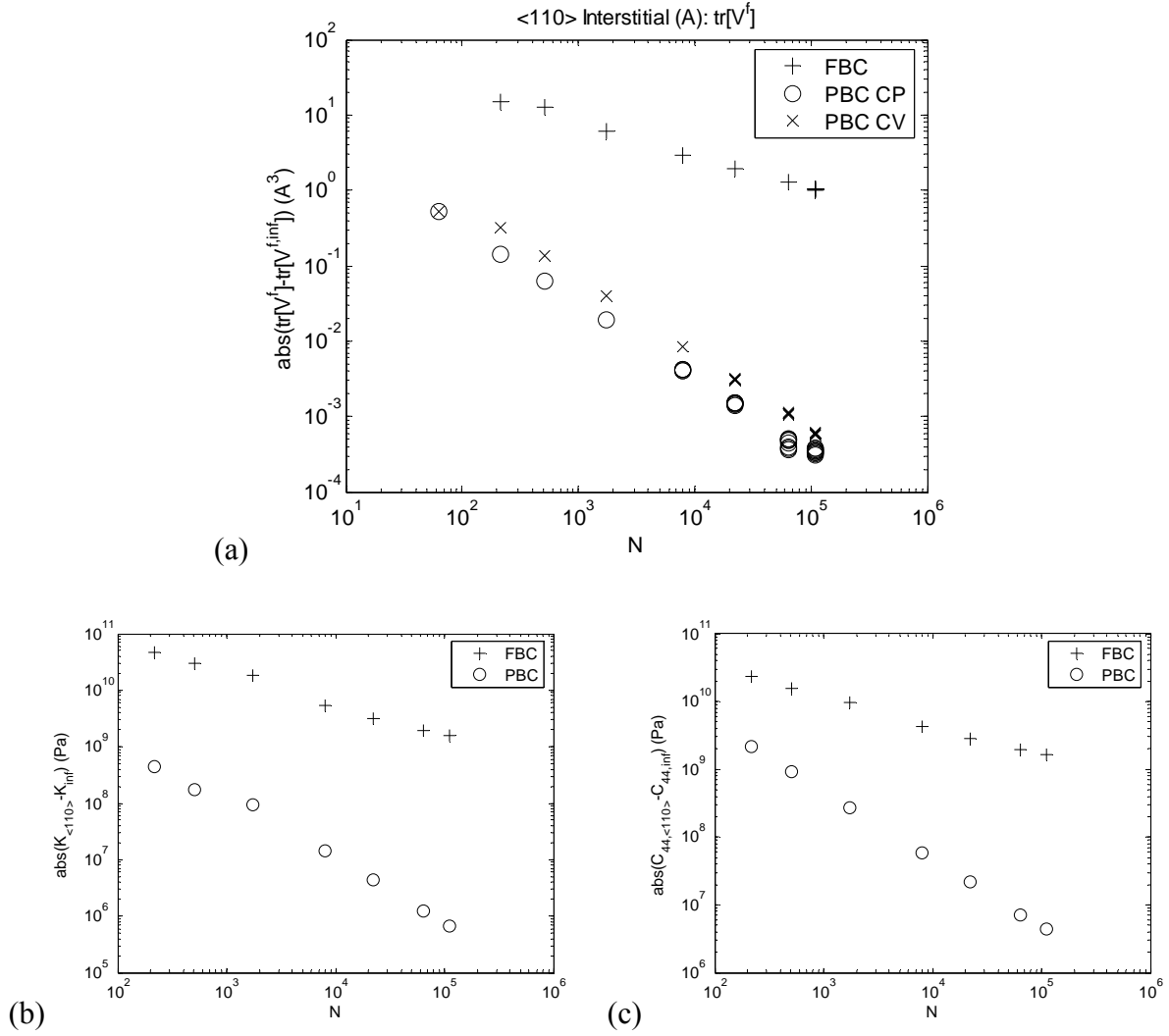

(a)

(b)                                    (c)

FIG. 5. The convergence of the formation volume (a) depends on the convergence of the moduli (b,c). The (b) bulk modulus and (c) shear modulus of systems with a <110> interstitial converge to the values ($K_{inf}$ = 1.0826 x $10^{11}$ Pa and $C_{44,inf}$ = 6.0256 x $10^{10}$ Pa) in defect-free Stillinger-Weber silicon in the large-size limit. The under coordination of surface atoms makes the convergence much slower in the free boundary condition case (FBC) than in the periodic boundary case with either constant pressure (PBC CP) or constant volume (PBC CV).

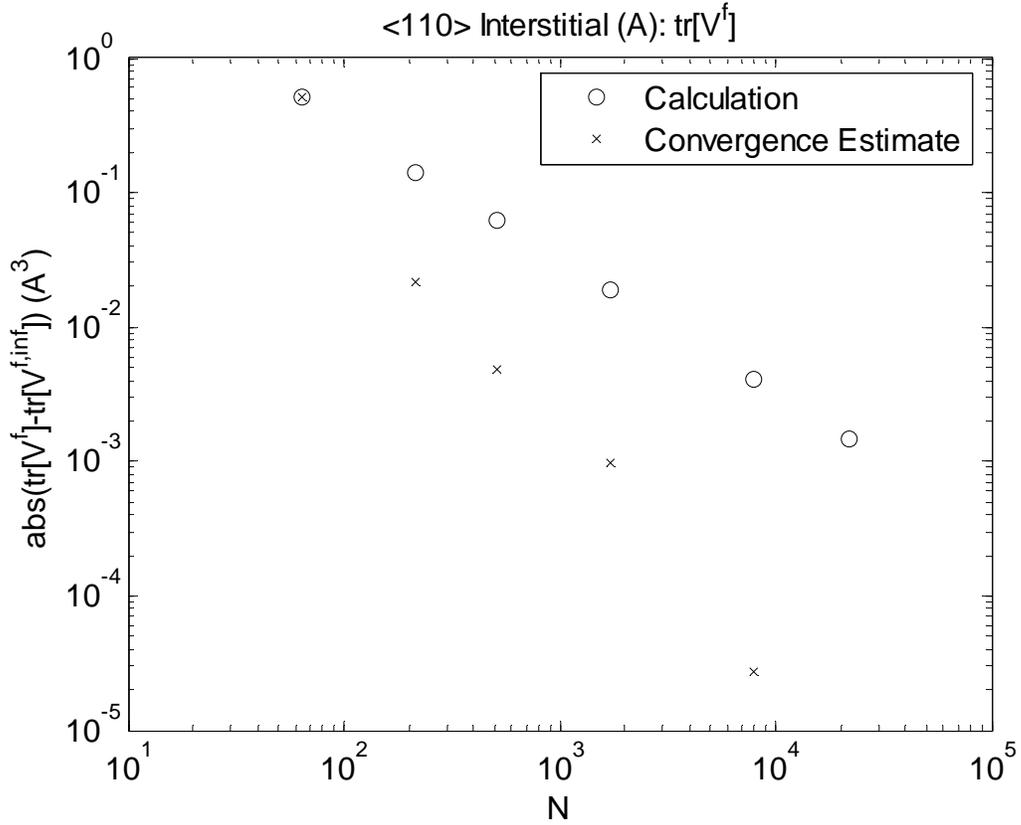

FIG. 6. The formation volume converges with error decreasing as $1/N$, the system size, shown above with (o). Knowing this convergence rate we can estimate a converged value from the two largest systems calculated using Eq. (19). This estimate, shown above with (x), is as converged at $N = 1728$ as the $N = 21,952$ calculation.